\documentclass[]{aa}
\usepackage{epsfig}
\def\lappr{\lower 3pt\hbox{$\buildrel < \over \sim\;$}}
\def\llappr{\lower 3pt\hbox{$\buildrel > \over \sim\;$}}
\begin{document}
    \title{Looking for obscured QSOs in the X--ray emitting ERO
    population}

   \author{P. Severgnini
          \inst{1}
	  \and
	 R. Della Ceca\inst{1}
	 \and
	 V. Braito\inst{1}
	 \and
	 P. Saracco\inst{1}
	 \and
	 M. Longhetti\inst{1}
	 \and
	 R. Bender\inst{2,5}
         \and
	 N. Drory\inst{3}
	 \and
	 G. Feulner\inst{2}
	 \and 
	 U. Hopp\inst{2,5}
	 \and 
	 F. Mannucci\inst{4}
	 \and
	 C. Maraston\inst{5}
	  }

   \offprints{P. Severgnini, e-mail: paola@brera.mi.astro.it}

   \institute{Osservatorio Astronomico di Brera, Via Brera 28, I-20121, 
     Milano, Italy\\
     \email{paola, rdc, braito, saracco, marcella@brera.mi.astro.it}
   \and
     Universit\"ats-Sternwarte M\"unchen, Scheiner Str. 1, 81679 M\"unchen, 
     Germany\\
     \email{bender, feulner, hopp@usm.uni-muenchen.de}
   \and
     University of Texas at Austin, Austin, Texas 78712\\
     \email{drory@astro.as.utexas.edu}
   \and
     IRA-CNR, Firenze, Italy\\
     \email{filippo@arcetri.astro.it}
    \and
      Max-Planck-Institut f\"ur extraterrestrische Physik,
      Giessenbachstra$\beta$e, D-85748 Garching, Germany \\
      \email{bender, hopp, maraston@mpe.mpg.de}
      }

   \date{}

   \abstract{
We present {\it XMM-Newton} data
centered on one of the MUNICS Near Infrared Cluster Survey fields (S2F1)
and we discuss the X-ray properties of the 6 X-ray emitting EROs found.
For one of them we have already obtained the redshift using
near--infrared spectroscopic data,
while for the remaining 5 EROs the analysis is based
on photometric redshifts.
We find evidence for the presence of an X--ray obscured QSO in at least
5 out of the 6 X-ray emitting EROs.
For these 5 objects we derive intrinsic (2--10 keV) 
luminosities in excess of 10$^{44}$ erg s$^{-1}$
and intrinsic column densities higher than 10$^{22}$ cm$^{-2}$.
These values have been obtained through a basic X-ray spectral
analysis for the three brightest sources and through
the analysis of the hardness ratios  for the remaining two.
All of these 5 X--ray emitting EROs appear extended in the 
optical/near--infrared bands
indicating that the host galaxy emission dominates at these wavelengths.
This suggests that the hosted AGNs are likely to be absorbed
also in the optical/near--infrared bands: i.e. they are likely 
X--ray obscured possible type~2 QSOs.
For the remaining ERO the presence of an AGN is suggested both by its
high 0.5-2 keV luminosity (L$_{0.5-2\rm keV}\sim10^{43}$ erg s$^{-1}$) 
and by its X-ray-to-optical flux ratio. In this case 
the quality of the present data prevents us from placing 
firm constraints on the AGN type hosted.
Finally, the near-IR spectrum obtained for one of the 6 EROs
classifies the host galaxy as an elliptical at $z\simeq1.7$ with a stellar
mass well in excess of 10$^{11}$ M$_\odot$. This result corroborates the
possible link between the QSO activity and the formation of massive spheroids.

   }

	\titlerunning{Looking absorbed QSOs in X--ray emitting EROs}
   \maketitle
%

\section{Introduction}

As shown by {\it Chandra} and {\it XMM-Newton} observations, a sizable fraction of the 2--10 keV 
selected sources are associated with Extremely Red  Objects (EROs,
R-K$>$5). In particular, about 15\% of the 2--10 keV sources down to a flux
limit  of 10$^{-15}$ erg cm$^{-2}$ s$^{-1}$ are EROs (e.g. Rosati et
al. \cite{Rosati}; Mainieri et al. \cite{Mainieri}).
This already significant fraction should be 
considered as a lower limit since several X--ray sources are not yet 
covered by deep K--band observations. Unfortunately, 
good--quality optical and/or X--ray spectra are available only for a
few X--ray emitting EROs,  preventing us from
firmly establishing their physical nature. 
The X--ray and optical/near--IR photometric properties,
the only information available for the large majority of the X--ray emitting
EROs studied so far, strongly suggest that the bulk of this
population is composed of obscured AGNs at least at the brightest
(F$_X>$10$^{-15}$ erg cm$^{-2}$ s$^{-1}$) X--ray fluxes
(e.g Alexander et al. \cite{Alexander02}, \cite{Alexander03}; 
Barger et al. \cite{Barger02}, \cite{Barger03}; Stevens et al. \cite{Stevens};
Szokoly et al. \cite{Szokoly}, Brusa et al. \cite{Brusa04}). 
Although dusty starbursts are also observed as EROs (e.g. Brusa et al. 
\cite{Brusa02}),
their fraction is not high enough to 
dominate the population of X--ray emitting EROs at these X--ray fluxes.
Indeed, the large majority of X--ray emitting EROs have measured 
X--ray--to--optical flux ratios  typical of AGN (i.e.
F$_X$/F$_{opt}>$0.1) and evidence of obscuration comes from both their hard
X--ray colors and their extremely  red optical--to--near--infrared colors.
In particular,  hard X--ray colors are generally associated with a large
amount of circumnuclear gas (N$_H$\footnote{As usually done in the 
recent literature, we use here N$_H$=10$^{22}$ cm$^{-2}$ as the
dividing N$_H$ value between X--ray unabsorbed and absorbed
sources.}$>$10$^{22}$ cm$^{-2}$) which blocks the
softer X--ray emission, while red optical colors indicate that large amounts
of dust are absorbing the typical intrinsic blue continuum of the AGN
(Elvis et al. \cite{Elvis}, Risaliti \& Elvis \cite{Risaliti}).
The few cases for which  either X--ray and/or optical 
spectra are available seem to confirm the presence of X--ray and/or optical 
obscuration (e.g. Lehmann et al. \cite{Lehmann}; Vignali et al. \cite{Vignali}; 
Mainieri et al. \cite{Mainieri}; Willott et al. \cite{Willott}; 
Roche et al. \cite{Roche}; Stevens et al. \cite{Stevens}; 
Mignoli et al. \cite{Mignoli}; Szokoly et al. \cite{Szokoly};
Brusa et al. \cite{Brusa}, \cite{Brusa04}) and, more important, indicate that most of 
the X--ray emitting EROs host high--luminosity 
AGNs (L$_X>$10$^{44}$ erg s$^{-1}$, i.e. QSOs).
These findings open the possibility that a significant fraction of
X--ray emitting EROs could host X--ray obscured and optically
absorbed (i.e type~2) QSO (for details on this class of sources see e.g. Norman et al. \cite{Norman};
Stern et al. \cite{Stern}; Della Ceca et al. \cite{Della Ceca03} 
and references therein). These sources are predicted in large numbers 
by the synthesis model
of the Cosmic X--ray Background (38\% and 16\% 
of Cosmic X--ray Background should be
due to QSO2 following the Gilli et al. \cite{Gilli} and
Ueda et al. \cite{Ueda} models, respectively).

To study the nature of X--ray emitting EROs,
we have obtained  two 
{\it XMM-Newton} observations
respectively on the  S2F1 (AO--2 period) and on the  S2F5 (AO--3 period) 
fields of the {\it MUnich Near-IR
Cluster Survey} (MUNICS, Drory et al. \cite{Drory}, Feulner et al.
\cite{Feulner}). These fields ($\sim160$ arcmin$^2$ each) are covered by
photometric observations in the B, V,  R, I, J and K' bands (Drory et al.
\cite{Drory}) down to limiting magnitudes of R$\simeq$24 mag and 
K'$\simeq$19.3 mag (50\% completeness limits for point sources). 
Thirty-six out of the 170 EROs present in these two 
fields have a
K' magnitude brighter than 18.5 and are already under investigation by
an ongoing spectroscopic survey performed by our group
(see  Saracco et al. \cite{Saracco1},  \cite{Saracco2}).

Here we discuss the X-ray properties
of 6 X-ray emitting EROs found in the S2F1 field. 
The {\it
XMM-Newton} data centered on the S2F5 field have not yet been made
available by the {\it European Space Agency}.
A comprehensive analysis of the whole sample of near-IR selected
EROs including the stacked analysis and the statistical X-ray
properties performed on both fields will be presented in a
forthcoming paper.

Throughout this paper we assume
H$_0$=70 km s$^{-1}$ Mpc$^{-1}$ and $\Omega_M$=0.3, $\Omega_{\Lambda}$=0.7.
All the magnitudes are in the Vega system.


\section{X-ray and NIR data} 

While the large majority of the X--ray emitting EROs studied so far are 
X--ray selected, here we deal with a sample of near--infrared selected
EROs for which follow--up X--ray observations have been obtained. 
Using the photometric data available from the
MUNICS catalog for the S2F1 field, we have selected $\sim$70 EROs down 
to a magnitude of K'$\simeq$19.3. This corresponds to
0.44$\pm$0.05 EROs/arcmin$^{2}$, in agreement with the surface density measured
in other surveys  at a comparable K--band magnitude  (e.g. 
Daddi et al. \cite{Daddi}). This field was observed by {\it XMM-Newton} on February
11, 2003 ({\it XMM-Newton} pointing: RA 03:06:41.56, Dec +00:01:00.4)  in full
frame mode and with the thin filter applied.  
In order to investigate the X--ray properties of our ERO sample,
we have considered the pipeline source
lists provided by the {\it XMM-Newton} SSC ({\it XMM-Newton  Survey Science
Center}, Watson et al. \cite{Watson01}) and  created using the SAS ({\it
Science Analysis System}) version  5.4.  Using the 8 X--ray brightest sources
in the field, we found that both for EPIC--pn and EPIC--MOS cameras the average
displacement between X--ray and optical (from the USNO catalog, 
Monet \cite{Monet}) positions is
of 3\arcsec$\pm$1.5\arcsec. 
We have  considered all the sources detected in at least
one of the {\it XMM-Newton} SSC energy bands\footnote{We refer to the following
six energy bands:  0.2--0.5, 0.5--2, 0.5--4.5, 2.0--4.5, 4.5--7.5, 7.5--12.0
keV; see http://xmmssc-www.star.le.ac.uk/ for details.} and we have  
cross-correlated their astrometrically corrected positions with the MUNICS 
K--band catalog containing all the near--infrared galaxies 
brighter than  K'$\sim$19.3 ($\sim$620) in the S2F1 field.
To this end, a conservative matching radius of 6\arcsec~has been used: 
this is the radius for which  more than 95\% of the {\it XMM--Newton}
sources in the SSC  catalog are
associated with USNO A.2 sources. 
We find that 6 EROs fall within 6\arcsec~of an X--ray source\footnote{ The
remaining near--infrared sources with an X--ray association and with R--K'$<$5 will
be discussed in a forthcoming paper.}.
For all of these 6 sources the offset between the X--ray and near--infrared position
is smaller than  4\arcsec.
Moreover,  no EROs
have been found with an X--ray association between 4\arcsec~ 
and 11\arcsec, implying that no NIR/X--ray
associations have been lost using the adopted  matching radius.
Using the probability of chance coincidence\footnote{$P=1-e^{-(\pi r^2\mu(<K'))}$, 
where $r$ is the matching radius used and $\mu(<K')$ is the cumulative 
surface density of EROs brighter 
than K' determined in the S2F1 field.}, we estimate that the expected 
number of 
spurious EROs/X--ray source associations is $<$0.8. 
The same result has been obtained
by cross--correlating the near--infrared positions
with fake X--ray coordinates (obtained by shifting the true 
X--ray positions by 1, 3 and 5 arcmin in different directions on the sky).

None of the 6 EROs discussed above has an optical
spectrum in the MUNICS catalog. All of them appear clearly extended
in the optical/near--infrared images even if the 
quality of the data is not good enough to allow a detailed
morphological analysis.

In four cases  
(S2F1\_71, S2F1\_493, S2F1\_507, S2F1\_714)\footnote{The
identification name used here is taken from the MUNICS catalog, see
http://capella.usm.uni-muenchen.de/~drory/munics/interactive/interactive.html}
no near--infrared sources,
in addition to the ERO, 
fall within 6\arcsec~of the X--ray centroids, thus yielding 
unambiguous matches. In two cases, 
more than one near--infrared source falls 
within the matching radius. Again, none of these galaxies
has an optical spectrum in the MUNICS catalog.
In particular, two field galaxies fall 
within 6\arcsec~of the X--ray source possibly associated with
the ERO S2F1\_443. 
The first galaxy (\#435, K'=18.4 mag) lies at 
5.4\arcsec~from the X--ray centroid. It appears extended in the 
optical/near--infrared
images and the SED is well fitted by an emission line galaxy at z$>$ 2. 
The second galaxy
(\#437, K'=18.3 mag), which is also an  SDSS source (SDSS J030632.1+000109.3, see
http://www.sdss.org/dr2/), lies 6\arcsec~away from the X--ray centroid. 
It appears extended with rather blue colors and its SED is well fitted by 
an emission--line galaxy at z=0.5.
Analogously,
one field galaxy (\#552, K'=18.9 mag) lies at a distance of 2\arcsec~from the
X--ray source associated with the ERO S2F1\_551.
As reported in the MUNICS catalog, the SED of the galaxy
\#552 is well fitted by a z=0.6 elliptical galaxy.
In order to disentangle these ambiguous matches, we used the 
chance coincidence probability and, whenever available, the radio detection.
In particular, we have estimated from our data the
cumulative surface density of field sources in K' band and we find 
P\footnote{In field galaxies, $\mu(<K')$ is the cumulative 
surface density of near--infrared sources brighter 
than K' determined in the S2F1 field (see footnote 4).}$_{field}$(435)=0.07, P$_{field}$(437)=0.09 (to be compared
with P$_{ERO}$(443)=9$\times10^{-5}$) and P$_{field}$(552)=0.02
(to be compared with P$_{ERO}$(551)=0.006).
The highly significant probabilities associated with EROs suggest the 
correct association with the X--ray sources.
For ERO S2F1\_443, the reality of being the X--ray emitting
source is also supported by the fact that it has 
a radio counterpart at less than 
2\arcsec~ from the XMM position; this ERO is the only one in our sample with a
NVSS ({\it NRAO VLA Sky Survey}, Condon et al. \cite{Condon}; 
f$_{(1.4\rm GHz)}$=26.1$\pm$0.9 mJy) and FIRST 
({\it Faint Images of the Radio Sky at Twenty-cm}, Becker et al. \cite{Becker}; 
f$_{(1.4\rm GHz)}$=24.38$\pm$0.14 mJy) counterpart. 
Taking into account the spectroscopic redshift of the object (see Table~1),
we obtain a L$_{(1.4\rm GHz)}\sim$5$\times$10$^{26}$ Watt/Hz,
typical of high power radiogalaxies (e.g. Sadler et al. \cite{Sadler}).

\begin{table}[t!]
\caption{NIR properties of the 6 X--ray emitting EROs.}
\label{NIR properties}
\small
\begin{center}
\begin{tabular}{lccl}
\hline
Name$^a$   & K'$^a$ & R-K'$^a$ &   z$_{phot}$$^a$ \\
           &	   &	      & 		 \\   				  
\hline
S2F1\_71  &  19.3  & $>$5     & 1.3		 \\
S2F1\_443 &  18.4  & $>$5.6   & 1.7$^b$	 \\
S2F1\_493 &  18.6  & 5.4      & 1.4		 \\
S2F1\_507 &  18.4  & 5.2      & 1.4		 \\
S2F1\_551 &  19.2  & $>$5     & 1.2		 \\
S2F1\_714 &  18.5  & $>$5.5   & 1.0		 \\
\hline
\hline
\end{tabular}
\end{center}
Notes:\\
$^a$ For more details on the near-infrared objects see 
the Interactive Catalog and Image Browser: \\
http://capella.usm.uni-muenchen.de/~drory/munics/interactive/\\
interactive.html.\\
The uncertainties on the photometric redshift are $\pm$0.2.\\
$^b$ For this ERO we have the spectroscopic redshift (see 
Sec.~4). For comparison, 
the photometric redshift reported in the MUNICS catalog is
z$_{phot}$=1.5.\\
\end{table}

\begin{table*}[t!]
\caption{NIR and X--ray cross--correlation results.}
\label{NIR and X--ray distance}
\small
\begin{center}
\begin{tabular}{lccclc}
\hline
Name   & EPIC Instruments&  $\Delta$(NIR-X)$^a$ &  P(ERO)	      & ML$_{pn}$$^b$ & Net counts$^c$\\   
           &                 &  [arcsec]	    &		      & &  [0.2-10 keV]\\						     
\hline
S2F1\_71   & MOS1+MOS2+pn   &  0.9		   &  3$\times$10$^{-4}$ & 106$^d$ & 193\\
S2F1\_443  & MOS1	    &  0.8		   &  9$\times$10$^{-5}$ & 41$^d$ & 35$^g$\\
S2F1\_493  &  pn	    &  3.8		   &  3$\times$10$^{-3}$ & 44$^e$ & 24$^h$\\
S2F1\_507  & MOS1+MOS2+pn   &  1.65		   &  4$\times$10$^{-4}$ & 165$^e$ & 181\\
S2F1\_551  & MOS2+pn	    &  4.0		   &  6$\times$10$^{-3}$ & 48$^d$ & 70\\
S2F1\_714  & pn	            &  2.2		   &  9$\times$10$^{-4}$ & 12$^f$ & 31$^i$\\
\hline
\hline
\end{tabular}
\end{center}
Notes:\\
$^a$ If the X--ray source has been detected 
with more than one instrument, 
the $\Delta$(NIR-X) is the mean distance weighted by the positional errors
associated with each instrument.\\
$^b$ From the {\it XMM-Newton} SSC source list. 
Full details about the source likelihood parameters can be found in 
http://xmmssc-www.star.le.ac.uk.\\
$^c$ Total net counts (calculated using XSPECv11.2, FTOOLSv5.0, see Sect. 3.1 and 3.2) 
in the 0.2--10 keV energy band from all the instruments reported in column~2.\\
$^d$ Source Likelihood in the 0.5--4.5 keV range (MOS1 camera).\\
$^e$ Source Likelihood in the 0.5--2.0 keV range (pn camera).\\
$^f$ Source Likelihood in the 4.5--7.5 keV range (pn camera).\\
$^g$ Since this source has been detected only in the 0.5--4.5 keV band, the net counts reported 
refer only to this band.\\
$^h$ Since this source has been detected only in the 0.5--2.0 keV band, the net counts reported 
refer only to this band.\\
$^i$ Since this source has been detected only in the 2--4.5 keV and 4.5--7.5 keV bands, 
the net counts reported  refer only to the 2-7.5 keV band.\\
\end{table*}

Thus, we find 6 firm X-ray emitting EROs.
The near--infrared properties of the 6 EROs
(K'--band magnitude and R-K' color) are reported in Table~1
(column~2 and 3) along with their redshift (column~4).
For all but one (S2F1\_443) we report the photometric redshift
taken from the MUNICS catalog.
The method used to obtain the MUNICS photometric redshifts and their 
accuracy
are discussed in Drory et al. (\cite{Drory03}).
As discussed below, AGNs are likely hosted
in these EROs. However, since these 6 EROs 
appear extended both in the
optical and near--infrared images, 
the host galaxy emission dominates at these wavelengths. For this reason 
the photometric redshift gives a reliable 
estimate of their redshift.
For ERO S2F1\_443 we report the spectroscopic redshift derived
by  a low-resolution near-IR spectrum obtained as a part of 
an ongoing spectroscopic follow-up program aimed at observing
the whole sample of bright (K'$<18.5$)
EROs in the field (Saracco et al. \cite{Saracco1}, \cite{Saracco2}).
We present this spectrum and the relevant analysis in Sec.~4.   

In Table~2 we report for each ERO
the EPIC cameras with which they have been detected (column~2),
the distance between the near--infrared and  the X--ray positions (column~3)
and the probability of finding the  ERO within this distance by chance 
(column~4).
The fifth column lists the maximum likelihood in the pn--image (ML$_{pn}$) 
for all the
EROs except for S2F1\_443 which lies on a pn gap;
for this source, the ML reported in the table refers to the MOS1 camera.
The last column reports net counts.
\section{X--ray analysis}
\begin{figure}[t!]
{\centerline{\epsfig{file=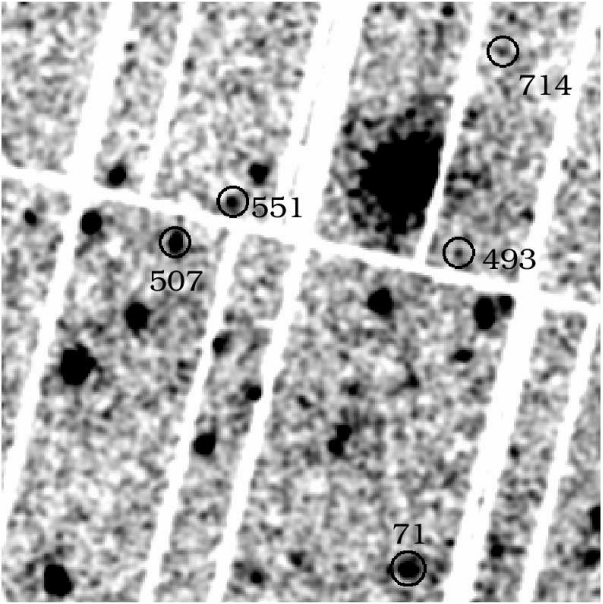, width=8cm, height=8cm, angle=0}}}
\caption{
Central part ($\sim12\arcmin\times12\arcmin$) of the 0.5--10 keV pn--image of the S2F1 field.
The open circles mark five of the six X--ray emitting EROs found in this field.
The sixth one (S2F1\_443) lies on a gap between the pn CCDs and it has only
a MOS1 detection.
}
\label{}
\end{figure}

As already discussed in Sect.~2, the {\it XMM-Newton} data of the S2F1 field 
have been  processed using tasks from the SAS version 5.4 and have 
been analyzed using
standard software packages (XSPECv11.2, FTOOLSv5.0, PIMMSv3.4). In order to
perform the X--ray analysis of our 6 EROs, the event files released
from the standard pipeline have been filtered for high background time
intervals. The net exposure times, after data cleaning, 
are $\sim$41 ksec, $\sim$44 ksec 
and $\sim$46 ksec for pn, MOS1 and MOS2 respectively.
We have used the latest calibration files released by the EPIC team to 
create response matrices that include also the correction for the 
effective area at the source position in the detector (SAS {\it arfgen} 
and {\it rmfgen} tasks have been used). All the 6 X--ray emitting 
EROs have an off--axis angle smaller than 6 arcmin. The central part 
($\sim$12\arcmin$\times$12\arcmin) of the 0.5--10 keV 
pn--image is shown in Figure~1. 
Five out of the six X--ray emitting EROs found in this field are marked with
a circle, while the remaining one (S2F1\_443, detected in the MOS1 camera) 
is not shown since it lies on a pn gap.
At the spatial resolution of the {\it XMM-Newton} EPIC instruments and
with the present statistics,
all the EROs appear to be X--ray point--like.

\subsection{EROs 71, 507, 551: X--ray spectral analysis}
For 3 EROs (71, 507, 551),
the number of net X--ray counts is good enough to allow a basic 
X--ray spectral analysis.
In order to maximize the signal--to--noise ratio,
the X--ray spectra have been extracted 
using circular regions of $\sim$20\arcsec~of radius centered on the X--ray
position. The background spectra have been extracted 
from source--free circular regions close to the object with an
area about 4 times larger.
While EROs S2F1\_71 and S2F1\_507 have been detected in all the 3 EPIC
instruments, ERO S2F1\_551 has not been detected in the MOS1 camera and
only barely detected in the MOS2 camera; therefore 
for this latter object only pn data have been used for the
spectral analysis. 
For the first two EROs
the MOS1 and MOS2 data have been combined to maximize the statistics. 
Finally, MOS and pn spectra 
have been binned in order to have at least 15 counts per 
energy channel and  are fitted simultaneously in the 0.5--10 keV band 
leaving the relative normalizations free to vary.
All the errors in the  following  are reported at the 90\%
confidence level for one interesting parameter ($\Delta\chi^2$=2.71, 
Avni \cite{Avni}).

In the fitting procedure, the appropriate Galactic hydrogen column density
along the line of sight  (6.85$\times$10$^{20}$ cm$^{-2}$, 
Dickey \& Lockman \cite{Dickey}) has been taken 
into account.
We find that the three X--ray spectra are well described
by a single absorbed (N$_H>10^{22}$ cm$^{-2}$) 
power--law model at the z$_{phot}$ of each source.
\begin{figure}[t!]
{\centerline{\epsfig{file=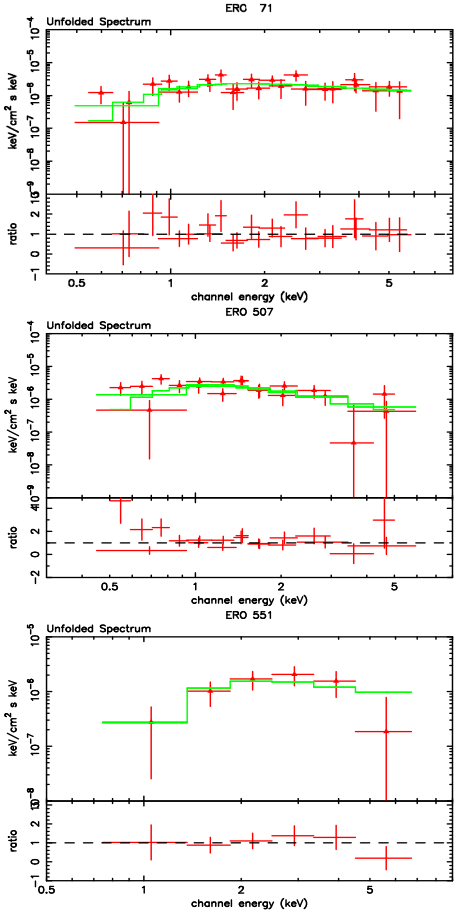,  width=4cm, height=6cm, angle=-90}}}
{\centerline{\epsfig{file=ero507_qdp.ps, width=4cm, height=6cm, angle=-90}}}
{\centerline{\epsfig{file=ero551_qdp.ps, width=4cm, height=6cm, angle=-90}}}
\caption{{\it XMM-Newton} spectra in energy units
(solid points) along with the best--fit models (continuous lines) and
the ratios between data and the best--fit model values (lower panels) as a
function of energy  for the 3 EROs reported in Table~3.}
\label{}
\end{figure}
Due to the X--ray statistics of ERO S2F1\_551 
($\sim$60 net counts in the pn image), 
the intrinsic photon index for this ERO is fixed to 1.9, the typical value
for unabsorbed AGN (Nandra et al. \cite{Nandra}; 
Caccianiga et al. \cite{Caccianiga}).

\begin{table*}[t!]
\caption{Best fit parameters for a single rest--frame absorbed power--law model.}
\begin{center}
\begin{tabular}{lcccccccccc}
\hline 
ID&  $\Gamma$ & $N_{\rm H}$$^a$ & $\chi^2/dof$ & F$_{(2-10\rm keV)}$$^b$ & L$_{(2-10\rm keV)}$$^c$ \\
& & $[10^{22}$ cm$^{-2}$] & & [10$^{-15}$ erg cm$^{-2}$ s$^{-1}$] & [10$^{44}$ erg s$^{-1}$] \\
\hline
S2F1\_71$^d$   & 1.47$_{-0.41}^{+0.78}$ & 2.2$_{-1.8}^{+4.4}$ &  16/22 & 19.4$\pm$3.6 & 1.4$_{-0.5}^{+0.7}$ \\
S2F1\_507$^d$  & 2.58$_{-0.44}^{+1.07}$ & 2.3$_{-1.8}^{+3.1}$ & 21.4/16 &  6.6$\pm$1.2 & 1.3$_{-0.5}^{+0.2}$\\
S2F1\_551  & 1.9(fixed) & 9.6$_{-5.9}^{+11}$ & 4.6/5 & 12.4$\pm$2.8  & 1.1$_{-0.4}^{+0.6}$\\
\hline
\hline
\end{tabular}
\end{center}
$^a$ Intrinsic column density.\\
$^b$ Fluxes are corrected only for the Galactic absorption.\\
$^c$ Luminosities are corrected both for the Galactic and 
intrinsic absorption.\\
$^d$ By using a fixed photon
index of 1.9 also for EROs S2F1\_71 and S2F1\_507
we obtain results similar to those reported in this table
except for the intrinsic absorbing column density
of S2F1\_507 which gives a best fit value
of N$_H\sim$0.8$\times$10$^{22}$ cm$^{-2}$
with a very large error bar.
\end{table*}
The best fit unfolded spectra and residuals are shown in Figure~2 while
the relevant best fit parameters,
quoted in the rest frame, are
summarized in Table~3 along with the 2--10 keV fluxes and the 
intrinsic luminosities. 
The errors reported in the table also include 
the uncertainty on the photometric redshifts ($\pm$0.2).
A pure thermal component is rejected for all the sources at more than 
97\% confidence level and the addition
of a thermal component to the power--law model is not statistically
required.
These results are in agreement with the
point--like appearance of the X--ray emission and with
the lack of evident massive clusters in the optical/near--infrared images.
As reported in Table~3, the intrinsic column densities and the 2--10 keV
luminosities, even if not extreme, are indicative of the presence of 
obscured (N$_H>10^{22}$ cm$^{-2}$) and high--luminosity 
(L$_{2-10\rm keV}>10^{44}$ erg s$^{-1}$) AGNs, i.e. X--ray obscured QSO.
All of them appear
extended in the optical and near-IR images, 
i.e. in these bands we do not observe
the  strong nuclear enhancement due to the QSO and the emission is 
dominated by the stellar continuum of the host galaxy. 
These facts suggest that the QSOs are likely to
be strongly absorbed also in the  optical/near--infrared domain, i.e.
they are X--ray absorbed type~2 QSO candidates.

\subsection{EROs 443, 493, 714: Broad-band properties} 

The remaining 3 EROs (443, 493, 714)  are detected only in one of the EPIC
cameras (see Table~2) and  the X--ray statistics are not good enough to obtain
a reliable X--ray spectrum. In particular, the ERO S2F1\_443 ($\sim$35 net
counts in the 0.5--4.5 keV band) is detected only in the MOS1 while  the 
S2F1\_493 ($\sim$24 net counts in the  0.5--2 keV) and  ERO S2F1\_714 ($\sim$31
net counts in the 2--7.5, in particular  $\sim$11 net counts in the 2--4.5 keV
plus $\sim$20 net counts in the 4.5--7.5 keV)  are detected 
only in the pn.  By performing a
visual inspection of all the 3 EPIC cameras, we find that, while S2F1\_493 and 
S2F1\_714 are only visible in the image where they have been detected, 
S2F1\_443 is also barely visible in the MOS2 image. In order to increase the
statistics of this source  we have used MOS1+MOS2 data thus accumulating
$\sim$50 net counts in total. Later, we will use these statistics to derive
the basic X--ray properties of S2F1\_443.

Some indications about the origin of the X--ray emission of these 3 EROs have
been derived by calculating their hardness ratios\footnote{HR is defined as
follow:  HR=$\frac{CR(2.0-4.5 keV)-CR(0.5-2.0 keV)} {CR(2.0-4.5
keV)+CR(0.5-2.0 keV)}$ where CR are the PSF and vignetting--corrected count
rates in the (0.5--2.0 keV) and in the (2.0--4.5 keV) bands.} (hereafter HR).
While the value of the HR ($\sim$--0.9) derived for S2F1\_493  is not a
discriminant of the X--ray emission origin, both S2F1\_443
(HR$_{443}\sim$--0.3) and S2F1\_714 (HR$_{714}\sim$1) have hardness ratios
typical of obscured  AGNs  (see Della Ceca et al. \cite{Dellaceca04}). By using
a simulated spectrum at the redshift of the  source and by assuming an
intrinsic photon index of 1.9, we have estimated that the intrinsic column
density needed to reproduce the hardness ratios of S2F1\_443 and S2F1\_714 are
consistent with values larger than 10$^{22}$ cm$^{-2}$ and of about 10$^{24}$
cm$^{-2}$, respectively. \\
\begin{figure}[t!]
{\centerline{\epsfig{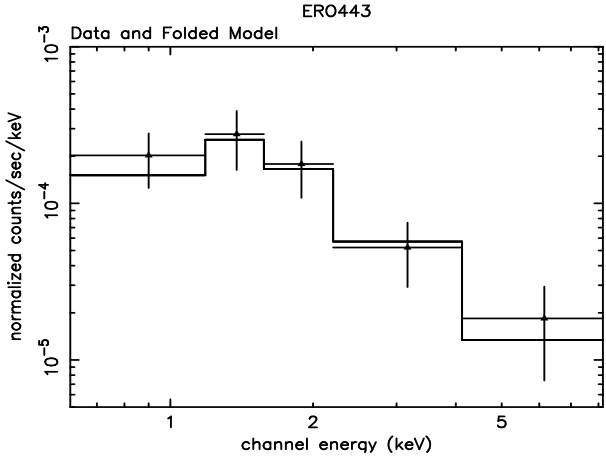}}} 
\caption{The
model (a rest-frame absorbed power--law with $\Gamma$=1.9 and
N$_H$=4$\times$10$^{22}$ cm$^{-2}$, continuous line)  derived from
the hardness
ratio is superimposed on the X--ray data of  the ERO S2F1\_443 (solid points).}
\label{} 
\end{figure}

Although the statistics of S2F1\_443 are not good enough to 
perform a complete X--ray spectral analysis, it allows us to compare the model used to
reproduce  the hardness ratio with the X--ray data. The good agreement between  model
and data is shown in Figure~3 where a rest-frame absorbed power--law model
($\Gamma$=1.9, N$_H$=3$\times$10$^{22}$ cm$^{-2}$) is superimposed on the
background--subtracted X--ray counts, binned in order to have at least 15 total
counts per energy channel. From this model we derive a Galactic corrected  flux of
F$_{(2-10\rm keV)}$=6.7$\pm$2.4$\times$10$^{-15}$ erg cm$^{-2}$ s$^{-1}$ and, by using
its spectroscopic redshift (1.7$\pm$0.05, see Table~1 and Sect.~4) we
estimate  an intrinsic luminosity of  L$_{(2-10\rm keV)}$=1.2$\pm$0.6$\times$10$^{44}$
erg s$^{-1}$ (the luminosity errors  take into
account also the redshift uncertainties), in good agreement with
its high radio luminosity (see Sect.~2).\\

\begin{figure*}[t!!!!!!!]
{\centerline{\epsfig{file=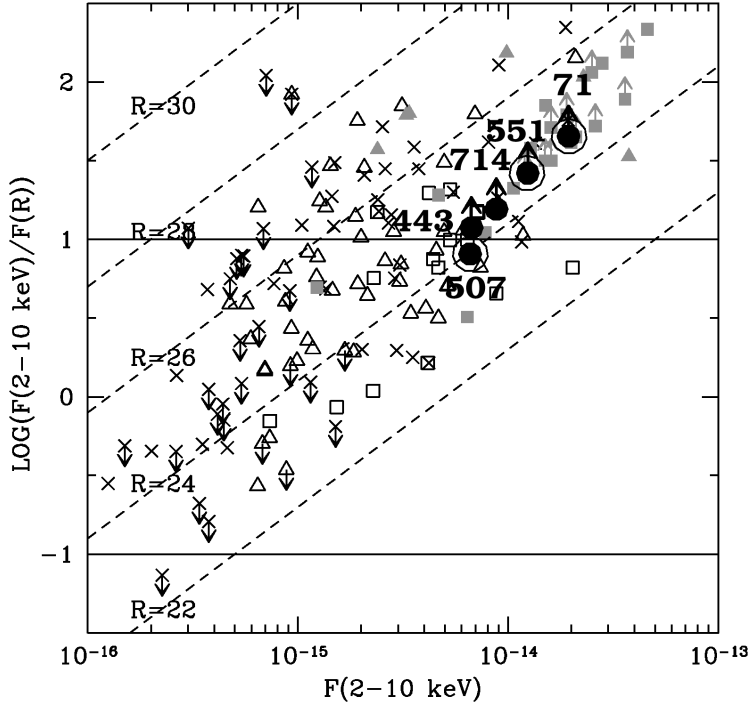, width=10cm, height=10cm, angle=0}}}
\caption{The logarithmic values of the (2--10 keV)--to--optical (R magnitudes)
flux ratios vs. the (2--10 keV) fluxes for five of our EROs (solid circles).
The X--ray fluxes have been corrected only for the Galactic absorption.
The 3 EROs for which the presence of an obscured and  high--luminosity AGN
has already been confirmed by the X--ray spectral analysis are marked with open
larger circles. EROs found in other surveys 
(Hellas2XMM, Mignoli et  al. \cite{Mignoli};
Lockman Hole, Mainieri et al. \cite{Mainieri}, Stevens et al.
\cite{Stevens}; CDFN \& CDFS, Alexander et al. \cite{Alexander02}, 
\cite{Alexander03}; Barger et al. \cite{Barger03}, Roche et al. \cite{Roche}, Szokoly et al. \cite{Szokoly}; 
ELAIS, Willott et al \cite{Willott}) are also plotted for comparison.
In particular: squares indicate the EROs for which there is firm
evidence of the presence of intrinsic obscuration on the basis of their 
X--ray and/or optical spectrum (solid symbols mark AGNs with
X--ray luminosity $>$10$^{44}$ erg s$^{-1}$; open squares mark AGNs with
X--ray luminosity $<$10$^{44}$ erg s$^{-1}$);
triangles mark the EROs for which X--ray obscuration is only 
suggested from the broad--band X--ray photometry, e.g. hardness ratios 
(solid symbols mark AGNs with
X--ray luminosity $>$10$^{44}$ erg s$^{-1}$; open triangles mark AGNs 
with
X--ray luminosity $<$10$^{44}$ erg s$^{-1}$); 
crosses indicate non--active galaxies (starburts and/or elliptical, 5 points
in total), EROs for which X--ray properties do not suggest
intrinsic obscuration (10 points in total) 
and EROs for which the information available are insufficient
to constrain their physical nature (47 points in total). 
Upper and lower limits of the (2--10 keV)--to--optical flux ratios
are marked with arrows.
The loci of constant magnitudes are marked as dashed lines.
The two solid lines define the region where 
luminous AGNs typically lie (Maccacaro et al. \cite{Maccacaro}; 
Schmidt et al. \cite{Schmidt}; Akiyama et al. \cite{Akiyama}; 
Lehmann et al. \cite{Lehmann01}).}
\label{}
\end{figure*}

In order to estimate the flux and the luminosity
for ERO S2F1\_714, we have
calculated the vignetting--corrected count rates in the 2--7.5 keV band
(the ERO has been detected only  in the 2--4.5 and 4.5--7.5 SSC band).
Using the measured count rates and 
assuming a power--law model with $\Gamma\sim$1.4 (similar to
that of the unresolved Cosmic X--ray background since the source seems to be
very hard) we obtain a Galactic corrected flux of 
F$_{(2-10\rm keV)}$=8.8$\pm$3.5$\times$10$^{-15}$ erg cm$^{-2}$ s$^{-1}$.
Taking into account the photometric redshift (1.0$\pm$0.2) and the 
intrinsic column density derived from the HR analysis (N$_H$ consistent with
10$^{24}$ cm$^{-2}$, see above) we estimate an intrinsic luminosity
L\footnote{The k--correction has been applied to the luminosities
assuming a power--law model with $\Gamma$=1.4.}$_{(2-10\rm keV)}>$10$^{44}$ erg s$^{-1}$.

In summary, we find evidence that 
both S2F1\_443 and S2F1\_714 are probably X--ray obscured 
AGNs of high 
luminosity ($>10^{44}$ erg s$^{-1}$).
This result is also supported by the
(2--10 keV)--to--optical flux ratios of these two EROs 
as a function of their (2--10 keV) fluxes (see Figure~4).
The two EROs discussed here are plotted with solid circles, while
the 3 EROs for which the presence of an obscured and high luminosity 
AGNs is already indicated by the X--ray spectral analysis (see Sec.~3.1) 
have been 
plotted with solid circles encircled by larger open circle. 
For comparison in Figure~4 we plot 
also other X--ray emitting EROs
taken from the literature (i.e. Hellas2XMM, Mignoli et  al. \cite{Mignoli};
Lockman Hole, Mainieri et al. \cite{Mainieri}, Stevens et al.
\cite{Stevens}; {\it Chandra Deep Field North}--CDFN, Vignali et al. \cite{Vignali}, 
Alexander et al. \cite{Alexander02}, \cite{Alexander03}, 
Barger et al. \cite{Barger03}; {\it Chandra Deep Field South}--CDFS, 
Roche et al. \cite{Roche}, Szokoly et al. \cite{Szokoly}; ELAIS, Willott et al 
\cite{Willott}).
Different symbols have been used to mark the
different EROs on the basis of the information available in the original papers.
In particular,  we have distinguished 
the EROs for which there is firm
evidence for the presence of intrinsic obscuration on the basis of their 
X--ray and/or optical spectrum (squares) from
the EROs for which X--ray obscuration is only suggested from the
broad--band X--ray photometry, i.e. hardness ratios (triangles). 
Non active galaxies, EROs for which X--ray properties do not suggest
intrinsic obscuration
and EROs for which the information available is insufficient
to constrain their physical nature are grouped together and indicated with 
crosses. More details on the
symbols are reported in the caption of the figure. 
A large fraction of the EROs 
having Log(F$_X$/F$_{opt})>$1 (at least 45 out of the 70) are 
obscured AGNs.
This is indeed what is expected 
on the basis of the observational results found so far by deep, medium and
bright X--ray surveys (Fiore et al. \cite{Fiore03}).
The fact that S2F1\_443 and S2F1\_714 lie in this region of the diagram
strongly supports the hypothesis derived from the hardness ratio 
analysis, i.e. that strong X--ray obscuration is likely present 
in these high luminosity AGNs.
Moreover, these 2 EROs also appear
extended in the optical and near-IR images suggesting that the AGNs are 
likely to be strongly absorbed also in the  optical/near--infrared domain.
All the X--ray and optical properties make 
the presence in these sources of X--ray obscured type~2 QSO likely.

As for ERO S2F1\_493, it is detected only in the 0.5--2 
keV energy band and it
has an HR value (HR$_{493}\sim$--0.9) 
which cannot exclude a starburst origin of
the X--ray emission observed. 
We have  calculated the
vignetting corrected count rates for this object 
in the detection band (0.5--2 keV) and
we have estimated the 0.5--2 keV flux corrected for the Galactic absorption
using both a power--law model with a photon index of 1.9 
(F$_{(0.5-2\rm keV)}$=0.9$\pm$0.4$\times$10$^{-15}$ erg cm$^{-2}$ s$^{-1}$)
and a Raymond-Smith model with KT=0.7 and Solar abundance 
(F$_{(0.5-2\rm keV)}$=0.8$\pm$0.4$\times$10$^{-15}$ erg cm$^{-2}$ s$^{-1}$). 
In both cases, taking into account the photometric redshift of the object
(1.4$\pm$0.2), we find a 0.5--2 keV   
luminosity of 
L$_{(0.5-2\rm keV)}\sim$10$^{43}$ erg s$^{-1}$. Even taking into
account the uncertainties, this value is more
typical of AGN and well exceeds that of starburst galaxies, typically lower
than 10$^{42}$ erg s$^{-1}$ (e.g. Persic et al. \cite{Persic}).
This makes it unlikely that the X--ray emission is dominated
by starburst activity unless 
an extremely high star formation rate, of the order of few 
1000 $M_{\odot}$ yr$^{-1}$ (Persic et al. \cite{Persic}), 
has occurred in this object.
On the other hand, with these star formation rates we would
expect a flux density at 1.4 GHz in excess of 1 mJy by using
the relation SFR($>5M_\odot$)=L$_{1.4GHz}$/(4$\times10^{28}$ [erg
s$^{-1}$ Hz$^{-1}$])  [M$_\odot$ yr$^{-1}$] (Condon \cite{Condon}; 
Persic et al. \cite{Persic}).
If this were the case, we would expect to detect 
this source in the FIRST survey
(the  rms at the source position is $\sim$0.144 mJy), 
while no radio counterpart is present.
On the contrary, its (0.5--2.0 keV)--to--optical flux ratio 
(F$_{(0.5-2 keV)}$/F$_{opt}$$\sim$1) is typical of unobscured AGN 
(Maccacaro et al. \cite{Maccacaro})
in agreement with the hardness ratio analysis. 
This fact, combined with its high 0.5--2 keV luminosity, 
 strongly supports the presence of an AGN.
The detection of this object only in the 0.5--2
keV energy band does not contradict the presence of an AGN with 
a photon index of 1.9. Indeed, we would not expect to detect 
such an AGN in the other {\it XMM-Newton}
SSC energy bands (see footnote 2) with the exposure time of our observation.
Moreover, the properties of this ERO do not exclude the presence of
a Compton--thick AGN where the intrinsic absorbing column density is well in
excess of 10$^{24}$ cm$^{-2}$ (see e.g. Ghisellini et al. \cite{Ghisellini}). 
If this were the case we would not directly observe the
X--ray emission produced by the central engine and we would 
probably measure
only the reflected fraction (at least a factor 10 lower
than the intrinsic X--ray emission). For this reason,
in the case of a Compton--thick AGN
the intrinsic luminosity of ERO 493
would be well in excess of 10$^{45}$ erg s$^{-1}$.
The hypothesis of high X--ray obscuration also 
would be in agreement with
its extended appearance in the optical/near--infrared images
 under the assumption of a good correlation between
optical and X--ray absorption (see Caccianiga et al. \cite{Caccianiga}).

\section{Near--infrared spectroscopy of ERO S2F1\_443}
\begin{figure}[t!]
{\centerline{\epsfig{file=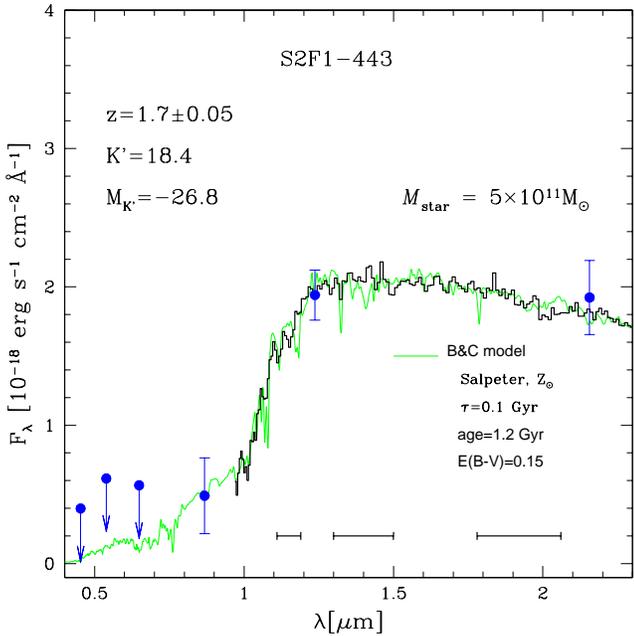, width=9cm, height=9cm, angle=0}}}
\caption{Smoothed Amici spectrum (thick black histogram) of the
ERO S2F1\_443. The filled symbols are the B, V, R, I ,J and K'
band photometric points from the MUNICS survey. The horizontal
bars mark the atmospheric windows with opacity larger than
80\%. A Bruzual \& Charlot (\cite{Bruzual}) template spectrum (thin grey line)
for 1.2 Gyr obtained with a Salpeter IMF at solar metallicity and a
decaying star formation time scale $\tau=0.1$ Gyr is superimposed
on the observed spectro-photometric data. 
The template is reddened by E(B-V)=0.15 mag.
The derived stellar mass
is 5$\times10^{11}$ M$_\odot$.
}
\label{}
\end{figure}
For the ERO S2F1\_443 we carried out near-IR spectroscopic
observations at the 3.6 Italian Telescopio Nazionale Galileo 
(TNG\footnote{see http://www.tng.iac.es}).
The observations (four hours of exposure) were performed in November 2003 
with the prism disperser Amici mounted on the near-IR camera NICS.
The prism allows us to cover the wavelength range 9000-24000 \AA~in 
a single shot with a constant  resolution of $\lambda/\Delta\lambda\simeq35$
(1.5'' width slit).
The Amici spectrum of the ERO S2F1\_443 is shown in Figure~5 (thick black histogram)
along with the optical and near-IR photometric points.
Given the low resolution of  Amici, this prism is best suited to describe 
the spectral shape of a source and to detect  strong continuum features
such as the 4000\AA~break, while it makes unfeasible the detection of 
emission/absorption lines for sources as faint as our EROs.
In particular, with the S/N reached in our observations, we expect
to detect only features with equivalent widths larger than $\sim150$ \AA.
Moreover, under the hypothesis that an obscured QSO is present,
the strongest lines expected in the Amici spectrum (e.g. [OIII] and
H$\alpha$) would lie in the
windows where the opacity is higher than 80\%.
For these reasons,
the lack of emission lines in the Amici 
spectrum of S2F1\_443 was expected even if it were hosting an obscured QSO.
On the other hand, the overall shape of the spectrum  unambiguously
classifies the host as an elliptical  galaxy. 
The Balmer break visible at $\lambda_{obs}\simeq1.05$ $\mu$m places this 
galaxy at $z=1.7\pm0.05$.
Given its apparent magnitude (K'=18.4), the resulting  rest-frame
(k-corrected) K-band absolute magnitude is M$_K=-26.8$,  i.e. 2.6 mag brighter
than local L$^*$ galaxies (M$^*_K$=-24.2, Cole et al. 2001).
Assuming that this galaxy evolves  passively  from $z\simeq1.7$ to $z=0$ 
($\Delta K\simeq1.2$ mag), i.e. the stellar mass of the galaxy is already fully 
assembled at $z\simeq1.7$, the resulting K-band luminosity of this elliptical 
at $z=0$ would be L$_{z=0}\simeq4$L$^*$.
Considering that local L$^*$ galaxies have stellar masses of the order
of 10$^{11}$ M$_\odot$, this implies that the elliptical S2F1\_443 has a 
stellar mass of the order of 4$\times10^{11}$ M$_\odot$.
This result, which is basically model--independent, shows that
this $z\simeq1.7$ elliptical possibly hosting  an  obscured QSO is
the high-z counterpart of the massive ellipticals populating the very bright 
end (L$_{z=0}\gg$L$^*$) of the local luminosity function.
A  comparison with the Bruzual \& Charlot (\cite{Bruzual}) models is also shown
in Figure~5. 
The synthetic spectrum (thin grey line) providing the 
best fit to the photometric and to the spectroscopic data has
been obtained with a Salpeter 
initial mass function with solar metallicity and an exponentially decaying  
star formation with a time scale $\tau=0.1$ Gyr. 
The template is reddened by E(B-V)=0.15 mag and it is 1.2 Gyr old, implying that
the bulk of the stars have to be formed at $z\sim3$ or higher
(Saracco et al. 2004, MNRAS submitted; Longhetti et al. 2004 in preparation).
The best--fitting model gives a stellar mass 
$\mathcal{M}_{star}=5\times10^{11}$ M$_\odot$ providing one of the most
massive ellipticals spectroscopically confirmed at these redshifts. 
The high mass and the  early-type nature of this galaxy hosting a QSO
and the similar results found in previous works (see e.g. Cowie et al.
\cite{Cowie}; Mignoli et al. 
\cite{Mignoli})
suggest a possible link between the  QSO activity and the end of the 
main episode of massive star formation in  spheroids 
(e.g. Granato et al. \cite{Granato}, \cite{Granato03}).
Higher resolution {\it near--infrared} spectroscopy and  
better statistics would be needed 
to fully address this  issue.

\section{Conclusions}
In this paper, new {\it XMM-Newton} data have been used to study and to 
discuss the nature of 6 X--ray emitting EROs found in the
MUNICS S2F1 field. For these 6 sources optical and near--infrared
photometry as well as photometric redshifts 
are available from the MUNICS catalog. For one of the 6 objects, ERO 
S2F1\_443, we have presented and analyzed the near--infrared low--resolution
spectrum, deriving  the spectroscopic redshift and unveiling 
the nature of the host galaxy.

Five (71, 443, 507, 551, 714) out of the 6 EROs discussed here have X--ray
properties matching those expected for X--ray obscured QSO.
For 3 of them (71, 507, 551) these properties have been derived by 
performing a basic X--ray spectral analysis, while for the other 2 
(443 and 714)
we used their X--ray broad--band photometry (i.e. X--ray emission, 
hardness ratio analysis, radio properties).
Due to the extended appearance of these 5 sources 
in the optical and near--infrared images, i.e. they are galaxy dominated
at these wavelengths, it is very likely that the QSOs are also strongly 
absorbed in the optical/near--infrared bands, i.e. they are also
type~2 QSO.
This is consistent with the fact that their 
(2--10 keV)--to--optical flux ratios vs. their 2--10 keV luminosities
are in agreement with the relation described by Fiore et
al. (\cite{Fiore03}) for type~2 AGNs.
In these 5 X--ray emitting EROs we find evidence for the presence 
of X--ray obscured type~2 QSO candidates.

For the remaining ERO, S2F1\_493, the presence of an AGN is strongly  suggested
by the high 0.5--2 keV luminosity and by the value of its  (0.5--2
keV)--to--optical flux ratio ($\sim$1). In this case the hypothesis 
of very high intrinsic
obscuration ($>$10$^{24}$ cm$^{-2}$) is not discarded. On the other hand, 
the quality of the data prevents us from placing reliable constraints on 
the AGN type hosted by this ERO. Due to its optical
X--ray broad--band photometry, if an obscured AGN is present, it would have an
X--ray luminosity in the QSO regime.

The only QSO host galaxy for which we have collected
spectroscopic observations so far is an elliptical galaxy
 at z$\simeq$1.7 with a stellar mass
well above 10$^{11} M_\odot$.
This result suggests
possible link between the  QSO activity and the formation of the bulk
of the stars in massive spheroids.


%

\begin{acknowledgements}
PS acknowledges financial support by the {\it 
Istituto Nazionale di Astrofisica} (INAF). 
This work has received partial financial support from ASI 
(I/R/037/01, I/R/062/02, I/R/071/02) and from the Italian Ministry of University and
the Scientific and Technological Research (MURST) through grant 
Cofin-03-02-23.
The MUNICS project was supported by the Deutsche
Forschungsgemeinschaft, Sonderforschungsbereich 375,
Astroteilchenphysik. We thank Alessandro Caccianiga, Tommaso Maccacaro, 
Christian Vignali and the anonymous referee for helpful comments.
\end{acknowledgements}

\end{document}